# Observing single $F_o F_1$-ATP synthase at work using an improved fluorescent protein mNeonGreen as FRET donor

Thomas Heitkamp[a], Gabriele Deckers-Hebestreit[b], Michael Börsch[a*]

[a] Single-Molecule Microscopy Group, Jena University Hospital, Friedrich Schiller University Jena, Nonnenplan 2 - 4, 07743 Jena, Germany
[b] Department of Microbiology, University of Osnabrück, Barbarastrasse 11, 49076 Osnabrück, Germany

## ABSTRACT

Adenosine triphosphate (ATP) is the universal chemical energy currency for cellular activities provided mainly by the membrane enzyme $F_o F_1$-ATP synthase in bacteria, chloroplasts and mitochondria. Synthesis of ATP is accompanied by subunit rotation within the enzyme. Over the past 15 years we have developed a variety of single-molecule FRET (smFRET) experiments to monitor catalytic action of individual bacterial enzymes *in vitro*. By specifically labeling rotating and static subunits within a single enzyme we were able to observe three-stepped rotation in the $F_1$ motor, ten-stepped rotation in the $F_o$ motor and transient elastic deformation of the connected rotor subunits. However, the spatial and temporal resolution of motor activities measured by smFRET were limited by the photophysics of the FRET fluorophores. Here we evaluate the novel FRET donor mNeonGreen as a fusion to $F_o F_1$-ATP synthase and compare it to the previously used fluorophore EGFP. Topics of this manuscript are the biochemical purification procedures and the activity measurements of the fully functional mutant enzyme.

## 1. INTRODUCTION

Since two decades, we are interested in measuring and understanding the mechanochemistry of the membrane enzyme $F_o F_1$-ATP synthase using single molecule spectroscopy[1-45] and biochemical assays[46-61]. To study the conformational dynamics we apply single-molecule Förster resonance energy transfer (smFRET) using a custom-designed confocal microscope. Reconstituted single enzymes in liposomes with ~120 nm diameter that diffuse freely in solution, generate a burst of photons during the transit through the confocal detection volume. Using a pair of fluorophores, i.e. the FRET donor and FRET acceptor, attached to a single $F_o F_1$-ATP synthase we established smFRET for the investigation of movements and distances between rotor and stator subunits in the $F_o F_1$-ATP synthase of *E. coli*.

A prerequisite for all smFRET experiments is the side-specific fluorescent labeling of the protein. Usually we labeled the $F_o$ and the $F_1$ domains separately by cysteine-reactive dyes and subsequently recombined both sub-complexes to obtain the functional FRET-labeled holoenzyme. However, to study the rotary motions within the $F_o$ motor (with one stator subunit *a* but 10 rotor subunits *c*, or for a three-color labeling scheme), cysteine mutants solely do not allow for specific labeling. An alternative label is a fluorescent protein, for example as a genetic fusion to subunit *a*. This has the advantage that all $F_o F_1$-ATP synthases are fluorescently tagged with an efficiency of nearly 100 % after fluorophore maturation. We also tried the SNAP-tag or the HALO-tag, but these fusion proteins could not be labeled after purification of $F_o F_1$-ATP synthase.

Here we evaluate a novel fluorescent protein mNeonGreen (Allele Biotechnology, USA)[62] as a fusion to the C terminus of subunit *a* of $F_o F_1$-ATP synthase. According to the supplier, mNeonGreen is a bright monomeric green/yellow fluorescent protein derived by structure-guided directed evolution from the tetrameric LanYFP from *Branchiostoma lanceolatum*. Excitation maximum at 506 nm with a high extinction coefficient, emission maximum at 517 nm and a quantum yield of 0.8 are reported resulting in an apparent 1.5 to 3-fold brightness increase compared to other GFP or YFP derivatives.

...................................................................................................................

*email: michael.boersch@med.uni-jena.de; http://www.single-molecule-microscopy.uniklinikum-jena.de

## 2. EXPERIMENTAL PROCEDURES

The $F_oF_1$-ATP synthase was genetically modified, expressed in *Escherichia coli* and purified from the plasma membrane. The procedures are given in detail below.

### 2.1 Construction of plasmid pACWU-BH1

To generate a fusion gene of a*tpB* (encoding subunit *a* of the $F_oF_1$-ATP synthase) and the gene encoding mNeonGreen (mNG), a *BamHI* restriction site (encoding a Gly-Ser spacer) and the gene encoding mNeonGreen were inserted by a two-step PCR overlap extension method prior to the stop codon of *atpB*. Three different PCR products were generated using (*i*) primer pair 5'-*GAATTCTCATGTTTGACAGC*-3' and 5'-<u>GTTATCCTCCTCGCCCTTGCTCAC CAT</u>**GGATCC***ATGATCTTCAGACGC*-3' with pBWU13.aGFP2 as template (pBWU13.aGFP2 is a pBWU13[63] derivative carrying an *atpB*-EGFP fusion gene with a *BamHI* restriction site as spacer), (*ii*) primers 5'-*GCGTCTGAAGATCAT***GGATCC**<u>ATGGTGAGCAAGGGCGAGGAGGATAAC</u>-3' and 5'-*CGTAGTAGTGTTG GTAAATTA*<u>CTTGTACAGCTCGTCCATGCCC</u>-3' with pNCS-mNeonGreen as template (Allele Biotechnology, USA), and (*iii*) primers 5'-<u>GGGCATGGACGAGCTGTACAAG</u>*TAATTTACCAACACTACTACG*-3' and 5'-*GGCTTCCGCTTTCGCTTTTTTCAGC*-3' with pBWU13 as template. Whereas the deoxynucleotides corresponding to the *atp* operon are in italics, those of mNeonGreen (mNG) gene are underlined and those of the spacer are in bold letters. Subsequently, the three PCR products were annealed and primers 5'-*GAATTCTCATGTTTGACAGC*-3' and 5'-*GGCTTCCGCTTTCGCTTTTTTCAGC*-3' used for the second amplification step. The resulting PCR product was restricted with *HindIII* and *PpuMI* yielding a 2214 bp fragment, which was introduced into correspondingly restricted pDC61 (kindly provided by S. D. Dunn; a pACWU1.2[64] derivative carrying an additional substitution of *c*E2C in the *atpE* gene). The fusion gene Φ(*atpB-mNG*)(Hyb) was verified by DNA sequencing through the ligation sites.

### 2.2 Expression strain and cell growth

The $F_oF_1$-*a*-mNeonGreen fusion construct was expressed in the *E. coli* strain RA1 *(F⁻ thi rpsL gal Δ(cyoABCDE) 456::KAN Δ(atpB-atpC)ilv:Tn10)*[65] which lacks most of the *atp* operon coding for the $F_oF_1$-ATP synthase. The pACWU-BH1 plasmid coding for the *atp* operon with the fusion of the *mNeonGreen* gene to the *atpB* gene was transformed into strain RA1. Cells were grown in a 10 l FerMac 320 fermenter (electrolab, UK) at 37°C, 800 rpm stirring and 4.5 l/min fresh air using a supplemented complex medium (0.5 g/l yeast extract, 1 g/l trypton, 17 mM NaCl, 10 mM Glucose, 107 mM $KH_2PO_4$, 71 mM KOH, 15 mM $(NH_4)_2SO_4$ and 4 μM uracil, 50 μM $H_3BO_3$, 1 μM $CoCl_2$, 1μM $MnCl_2$, 2 μM $ZnCl_2$, 10 μM $CaCl_2$, 3 μM $FeCl_2$, 0.5 mM $MgSO_4$, 0.5 mM arginine, 0.5 mM isoleucine, 0.7 mM valine, 4 μM thiamine and 0.4 μM 2,3-dihydroxybenzoic acid). When cell growth reached the late logarithmic phase ($OD_{600nm}$ = 0.8), cells were harvested by centrifugation in a Sorvall Evolution RC (Thermo Fisher Scientific, USA) at 10,000 x g and 4°C for 5 min. The cell pellets were flash frozen in liquid nitrogen and subsequently stored at -80°C.

### 2.3 Purification of $F_oF_1$-*a*-mNG ATP synthase

The fusion enzyme $F_oF_1$-*a*-mNeonGreen was essentially purified like the 'wildtype' protein[41]. However, low expression levels of the mutant construct and different equipment required modifications of the protocol. Unless otherwise noted, all purification steps and pH adjustments were performed at 4°C. Phenylmethylsulfonylfluoride (PMSF) was added freshly before use.

The frozen cells from two fermenter runs (32 g) were thawed slowly in 50 mM Tris-HCl pH 7.5 and then collected by centrifugation at 15,000 x g for 15 min. The cell pellet was resuspended in 150 ml of 50 mM MOPS pH 7.0, 175 mM KCl, 10 mM $MgCl_2$, 0.2 mM EGTA, 0.2 mM DTT and 0.1 mM PMSF (lysis buffer). A small amount of DNase I was added and cells were lysed by two passages through a cell homogenizer (PandaPlus 2000, GEA Niro Soavi, Italy) at 1000 bar. The partly lysed cells were cooled down to 4°C (ice water bath) between the two passages. The cell lysate was subsequently flash frozen by dropping into liquid nitrogen using a peristaltic pump P1 (Pharmacia, Sweden) at a flow rate of 5 ml/min. The frozen pearls were stored overnight at -80°C.

The combination of the *E. coli* strain RA1 with the particular cell homogenisator seemed to result in very small membrane vesicles difficult to centrifuge. This effect was heavily dependent on the amount of cells used. Therefore, the following procedure of cell membrane purification was performed twice with one half of the cell lysate. The cell lysate pearls were thawed slowly in a 10°C water bath, filled up to 150 ml with lysis buffer and supplemented with 0.25 mM PMSF. The cell debris was separated by centrifugation at 25,000 x g for 20 min (Sorvall Evolution RC, Thermo Fisher Scientific, USA). The supernatant was filled up to 200 ml with lysis buffer and membranes were collected by ultracentrifugation at 300,000 x g for 2 h using a 70 Ti rotor (Optima XP, Beckman Coulter, USA). The membrane pellets were resuspended in 150 ml 5 mM Tris-HCl pH 8.0, 5 mM MgCl$_2$, 0.2 mM EGTA, 0.2 mM DTT, 6 mM para-aminobenzamidine (PAB), 10% (v/v) glycerol and 0.1 mM PMSF using a soft marten hair paintbrush and centrifuged again at 300,000 x g for 1.5 h in the same rotor. This washing step was repeated once, thereby reducing the total volume to 100 ml. After centrifugation, the tubes were dried with a paper towel and the weight of the membrane pellets was determined. Finally, membrane pellets were resuspended in 12 ml 20 mM MES/Tricine-KOH pH 7.0, 5 mM MgCl$_2$, 2 mM DTT and 0.001% (w/v) PMSF (solubilization buffer), flash frozen by dropping into liquid nitrogen and stored overnight at -80°C.

Frozen membrane pearls were quickly thawed in a water bath and filled up with solubilization buffer. The membrane proteins were solubilized using 1.75 % (w/v) n-dodecyl β-D-maltoside (DDM; Glycon, Germany) in 6 ml solubilization buffer per g DDM. The final volume of the solubilization mixture was 10 ml per g membrane. After the dropwise addition of the DDM solution, the mixture was incubated on ice for 2 h with gentle stirring. Subsequently, the solubilized proteins were separated by ultracentrifugation at 300,000 x g for 2 h. The supernatant containing the solubilized proteins was concentrated and lipids removed by a two-step (NH$_4$)$_2$SO$_4$ precipitation. In the first step, impurities were precipitated with 45% (v/v) ice-cold, saturated (NH$_4$)$_2$SO$_4$. After centrifugation at 25,000 x g for 15 min, the supernatant was mixed with ice-cold saturated (NH$_4$)$_2$SO$_4$ to a final concentration of 65% (v/v). Under these conditions, most of the remaining proteins including the F$_o$F$_1$-ATP synthase precipitated. The precipitate was supplemented with 0.001% (w/v) PMSF and stored overnight at 4°C.

After centrifugation at 25,000 x g for 15 min, the protein pellet was resolved in 2 ml 40 mM MOPS-KOH pH 7.5, 80 mM KCl, 4 mM MgCl$_2$, 2 mM DTT, 2% (w/v) sucrose, 10% (v/v) glycerol, 1% (w/v) DDM and 0.001% (w/v) PMSF. Small particulate matter was removed by filtering through a 2 μm cellulose acetate syringe filter (Roth, Germany). Subsequently, the filtrate was loaded on a self-packed XK16/100 Sephacryl S300 size exclusion column (GE-Healthcare, USA), which was equilibrated with the same buffer containing 0.1% (w/v) DDM (SEC buffer). The column was cooled to 10°C by a cooling jacket connected to a chiller FL601 (Julabo, Germany). Proteins were eluted using an Äkta PrimePlus FPLC system (GE-Healthcare, USA) at a flow rate of 0.6 ml/min. The eluate was collected in 4 ml fractions and light absorbance at 280 nm and buffer conductivity was continuously measured.

F$_o$F$_1$-ATP synthase-containing peak fractions were filtered as described above and then loaded separately on a Poros HQ 20 (4.6 x 100 mm) anion exchange column (Applied Biosystems, USA). The column was connected to a BioCad 60 HPLC system (Applied Biosystems, USA). Chromatography was performed at room temperature with a flow velocity of 3.5 ml/min. The buffers were blended directly by the machine using a double concentrated SEC buffer, H$_2$O and 3 M KCl. First, the column was equilibrated by 10 column volumes (CV) of SEC buffer. Then, the sample was automatically injected onto the column using an autoinjector system. After washing the column with 5 CV, the ATP Synthase was eluted by a linear gradient over 12 CV up to 0.81 M KCl. Strongly bound impurities were eluted by a steeper gradient over 6 CV up to 1.54 M KCl. During the run, buffer conductivity and absorbance at 280 nm were measured and the eluate was collected in 1 ml fractions.

The F$_o$F$_1$-ATP synthase-containing anion exchange peak fractions were pooled, supplemented with 0.001% (w/v) PMSF and subsequently mixed with ice-cold, saturated (NH$_4$)$_2$SO$_4$ to a final concentration of 65% (v/v). The precipitated protein was collected by centrifugation at 25,000 x g for 15 min. and resolved in 0.5 ml SEC buffer. After centrifugation of the resolved protein for 10 min at 20,000 x g, the supernatant was loaded on a second size exclusion column. The Tricorn Superose 6 10/300 GL column (GE Healthcare, USA) was connected to an Äkta PrimePlus FPLC system (GE-Healthcare, USA) and equilibrated with SEC buffer for 2 CV at room temperature. The proteins were eluted at a flow rate of 0.4 ml/min at room temperature and collected in 0.5 ml fractions. Light absorbance at 280 nm and conductivity were continuously measured. The peak fractions were shock frozen in liquid nitrogen in 250 μl aliquots in cryo-straws and stored at -80 °C.

### 2.4 Fluorescence labeling of the *c* subunit

Cysteine labeling reactions were carried out in the dark. Purified F$_o$F$_1$-*a*-mNeonGreen containing a cysteine at residue position 2 of the *c* subunit was first concentrated *via* precipitation with 65% (v/v) ice-cold, saturated (NH$_4$)$_2$SO$_4$. The precipitated enzyme was centrifuged for 30 min at 20,000 x g and the resulting pellet was resolved in 75 μl 50 mM MOPS-NaOH pH 7.0, 0.1 mM MgCl$_2$ and 0.1 % DDM (labeling buffer). To separate residual (NH$_4$)$_2$SO$_4$ and DTT, the resolved protein was subjected to desalting size exclusion chromatography. Therefore, a self-made PTFE-frit was pressed onto the bottom of a 1 ml disposable syringe (B. Braun, Germany) and the syringe was then filled to 5 mm below the rim with Sephadex G50 medium (GE Healthcare, USA), pre-swollen in water. A cut 200 µl pipette tip was pressed onto the top as a funnel. This column was equilibrated by gravity flow with 3 ml of labeling buffer and excess mobile phase was removed by centrifugation for 2 min with 250 x g at room temperature. After loading of the resolved protein, the column was put into a glass vial with a 0.5 ml Eppendorf reaction tube at the bottom and centrifuged for 2.5 min at 445 x g. The concentration of the desalted enzyme was determined by mNeonGreen absorption at 506 nm (Lambda 650, PerkinElmer, USA) using a molar extinction coefficient ($\varepsilon_{506nm}$) of 116,000 L·mol$^{-1}$·cm$^{-1}$. The cysteine in the $c$ subunit was labeled in three independent reactions with AlexaFluor 568 C$_5$-maleimide, AlexaFluor 647 C$_2$-maleimide (Thermo Fischer Scientific Inc., USA) or ATTO 647N maleimide (ATTO-TEC, Germany). Labeling was carried out on ice in a total volume of 100 µl using a protein concentration of 6 µM and a molar ratio of F$_o$F$_1$ to dye of 1:1. This corresponds to a molar ratio of $c$ subunit to dye of 1:10. A five-fold molar excess of Tris(2-carboxyethyl)phosphine (TCEP) to subunit $c$ was used to convert the cysteines to the reactive reduced form. The reaction was stopped after 15 minutes by three consecutive Sephadex G50 medium columns as described above. The labeling efficiency was determined by measuring the absorption of mNeonGreen at 506 nm and the corresponding acceptor dye absorption at its maximum (AlexaFluor 568: $\varepsilon_{580nm}$ = 92,000 L·mol$^{-1}$·cm$^{-1}$; AlexaFluor 647: $\varepsilon_{654nm}$ = 265,000 L·mol$^{-1}$·cm$^{-1}$; ATTO 647N: $\varepsilon_{650nm}$ = 150,000 L·mol$^{-1}$·cm$^{-1}$). Labeling specificity was checked by 12 % SDS-PAGE with subsequent fluorography. Finally, the labeled protein was adjusted to 10 % (v/v) glycerol, flash frozen as 5 µl aliquots in liquid N$_2$ and stored at -80°C.

## 2.5 Reconstitution of F$_o$F$_1$-ATP synthase

To measure synthesis activity or smFRET, the F$_o$F$_1$-ATP synthase was reconstituted into pre-formed liposomes. This was achieved by slightly modifying established protocols[66, 67]. Briefly, pre-formed liposomes were generated by evaporating chloroform solutions of phosphatidylcholine and phosphatidic acid (Lipoid GmbH, Germany) in a mass ratio of 19:1 and subsequently resolving the mixture to 18 g/l in 10 mM Tricine-NaOH pH 8.0, 0.1 mM EDTA, 0.5 mM DTT, 7.2 g/l cholic acid and 3.6 g/l desoxycholate. The solution was sonicated three times for 30 s on ice (Digital sonifier 250: microtip, level 3, constant duty cycle; Branson Ultrasonic SA, Switzerland) and dialyzed at 30°C for 5 h against the 4,000-fold volume of 10 mM Tricine-NaOH pH 8.0, 0.2 mM EDTA, 0.25 mM DTT and 2.5 mM MgCl$_2$ using a Diachema membrane type 10.14 MWC 5000 (Diachema AG, Switzerland). After dialysis, the liposomes with an approximate concentration of 16 g/l were flash frozen in liquid N$_2$ and stored in 500 µl cryo straws at -80°C.

FRET-labeled and unlabeled F$_o$F$_1$-$a$-mNeonGreen as well as 'wildtype' F$_o$F$_1$-ATP synthase were reconstituted into pre-formed liposomes as follows. In a 2 ml reaction tube, 200 µl of pre-formed liposomes, 1 µl of 1 M MgCl$_2$ and reconstitution buffer (20 mM Tricine-NaOH pH 8.0, 20 mM succinate, 50 mM NaCl and 0.6 mM KCl) were mixed with protein to yield 43.5 nM protein in 368 µl. Under vigorous mixing (Vortex Genie 2, Scientific Industries, Inc.) 32 µl of 10 % (v/v) Triton-X-100 were added to destabilize the liposome membranes. After 1 h incubation with gentle shaking, 128 mg of pre-treated BioBeads SM-2 (Biorad, USA)[68] were added to remove the detergent. After 1 hour, the proteoliposomes were separated from the BioBeads. An aliquot was used directly for ATP synthesis measurements. The remaining reconstituted enzyme was adjusted to 10 % (v/v) glycerol, flash frozen as 5 µl aliquots in liquid N$_2$ and stored at -80°C. The final protein concentration in the reconstituted sample was 40 nM.

## 2.6 Measurement of ATP hydrolysis

The continuous measurement of ATP hydrolysis was carried out in an coupled enzymatic assay in a spectrophotometer at 21°C similar as described[69, 70]. Solubilized enzyme (3-10 nM) was added to the reaction buffer (100 mM Tris-HCl pH 8.0, 25 mM KCl, 4 mM MgCl$_2$, 2.5 mM phosphoenolpyruvate, 18 units/ml pyruvate kinase, 16 units/ml lactate dehydrogenase, 2 mM ATP and 0.4 mM NADH). ATP hydrolysis was detected by the decrease of NADH absorbance.

The reaction buffer was pre-heated to 37°C, placed in the spectrophotometer and the baseline was recorded at 340 nm. After adding the solubilized F$_o$F$_1$-ATP synthase, a decrease of absorbance was recorded. The rate of ATP hydrolysis was calculated from the linear slope using an extinction coefficient of 6,220 L·mol$^{-1}$·cm$^{-1}$ for NADH. The activation of F$_o$F$_1$-ATP synthase was proven by adding 20 µl of a 30 % (v/v) N,N-dimethyl-n-dodecylamine N-oxide (LDAO) solution to the assay mixture, corresponding to a final concentration of 0.6 %.

## 2.7 Measurement of ATP synthesis

The rate of ATP synthesis was determined as described[67]. The increase of the ATP concentration was observed continuously *via* luciferin/luciferase luminescence in a luminometer (1250 luminometer, Bio-Orbit , Finland) using an ATP-monitoring kit (Roche, Switzerland). To 'energize' the proteoliposomes, an acid/base transition and an additional

K$^+$/valinomycin diffusion potential was generated. Therefore, the proteoliposomes were incubated in the acidic medium (buffer LI: 20 mM succinate-NaOH pH 4.7, 0.6 mM KOH, 5 mM NaH$_2$PO$_4$, 2.5 mM MgCl$_2$, 0.1 mM ADP and 20 µM valinomycin) and added to the basic medium (buffer LII: 200 mM Tricine-NaOH pH 8.8, 160 mM KOH, 5 mM NaH$_2$PO$_4$, 2.5 mM MgCl$_2$ and 0.1 mM ADP).

Measurements were carried out at 21°C as follows: 880 µl of the basic buffer were mixed with 20 µl ATP-monitoring kit in a polystyrene test tube, the mixture was placed in the luminometer and the baseline was recorded with a pen recorder (LKB 2210 Recorder, Pharmacia, Sweden). To 20 µl proteoliposomes (with a F$_o$F$_1$-ATP synthase concentration of 40 nM) 80 µl of acidic medium were added carefully. After incubation for 3 min, ATP synthesis was initiated by quick injection of the 100 µl proteoliposomes with a Hamilton syringe through a septum directly into the basic medium inside the luminometer. The increase of the ATP concentration was directly proportional to the gain of luminescence intensity. The measurement was stopped after the signal reached a constant level. The rate of ATP synthesis was calculated from the initial slope and by means of ATP calibration. Therefore, the luminescence of a mixture of 970 µl basic medium and 20 µl ATP-monitoring kit was measured before and after adding 10 µl ATP standard solution (concentration 10 µM). All values given are the arithmetic average of at least duplicate measurements plus standard deviation.

## 2.8 Other methods

Concentrations of unlabeled 'wildtype' enzyme were determined by absorbance at 280 nm using a molar extinction coefficient of 310,120 L·mol$^{-1}$·cm$^{-1}$. Sodium dodecyl sulfate polyacrylamide gel electrophoresis (SDS-PAGE) was carried out as described[71] with an acrylamide concentration of 12 %. Subsequently, gels were stained with coomassie-R250.

# 3. RESULTS

## 3.1 Bacterial growth with plasmid-encoded fluorescent F$_o$F$_1$-*a*-mNeonGreen ATP synthase

The fluorescent protein mNeonGreen was fused to the C-terminus of the membrane-integrated subunit *a* of the F$_o$ domain, separated by a Gly-Ser linker. The fusion gene was inserted into the operon *atpBEFHAGDC* generating plasmid pACWU-BH1 (Figure 1) which is a low copy number plasmid (pACYC177 derivative) and encodes the eight structural genes essential for a functional F$_o$F$_1$-ATP synthase. After transformation of the *Escherichia coli atp* deletion strain DK8 (*bglR thi1 rel1* HfrPO1 Δ*atpB-C ilv*::Tn*10* (Tet$^R$))[72] with plasmid pACWU-BH1, growth on minimal medium with succinate as sole carbon and energy source was determined to confirm the presence of a functional F$_o$F$_1$-ATP synthase. With succinate, a substrate of the citric acid cycle, efficient ATP synthesis is possible only *via* oxidative phosphorylation which couples the proton-motive force generated by respiration to ATP synthesis from ADP and inorganic phosphate by F$_o$F$_1$. Growth of strain DK8/pACWU-BH1 was comparable to DK8/pBWU13 carrying 'wildtype' *atp* genes, or DK8/pDC61 which is the direct parent of plasmid pACWU-BH1 (Figure 1).

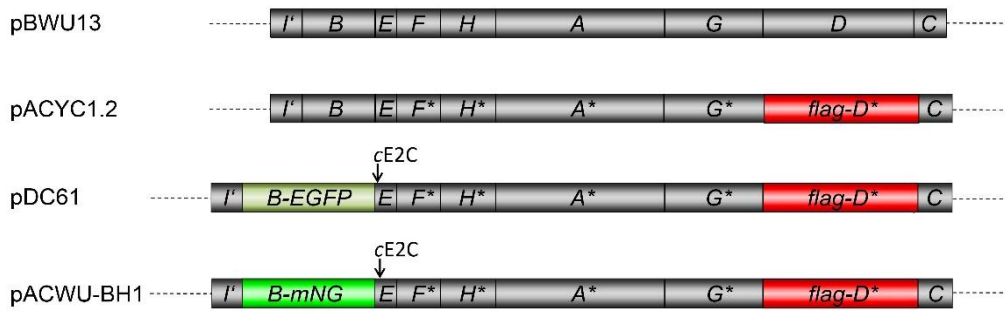

**Figure 1.** Generation of plasmid pACWU-BH1. The cartoon illustrates the changes in pACWU-BH1 in comparison with the parental wild-type *atp* operon (pBWU13) and intermediate plasmids pACYC1.2 and pDC61, respectively. *Grey*: *atp* genes, *green*: *atpB-mNG* (mNeonGreen) fusion gene with both genes linked via 6 additional nucleotides encoding a Gly-Ser spacer, *red*: *flag-atpD* codons for a FLAG-tag (DYKDDDDK) were inserted after the start codon of *atpD*, \*: codons encoding native cysteine residues were substituted by alanine codons (protein level), namely αC47A, αC90A, αC193A, αC243A, βC137A, γC87A, γC112A, δC65A, δC141A, *b*C21A (numbering always starts with f-Met at position 1). Arrow: substitution of codon 2 in the *atpE* gene resulting in the change *c*E2C (cysteine mutant in subunit *c*).

## 3.2 Expression and purification of $F_oF_1$-$a$-mNeonGreen ATP synthase

The expression level of the $F_oF_1$-ATP synthase in the bacterial membrane is crucial for efficient purification. Earlier attempts to purify enzyme mutants with significant lower expression levels than 'wildtype' always resulted in a markedly reduced purity. Therefore, samples of the membrane fractions of $F_oF_1$-$a$-mNeonGreen were compared to the expression of 'wildtype' $F_oF_1$ and a previous mutant $F_oF_1$-$a$-EGFP (Figure 2). The new mNeonGreen fusion construct showed the same reduced expression level as the EGFP fusion as indicated by the reduced bands of the α and β subunits.

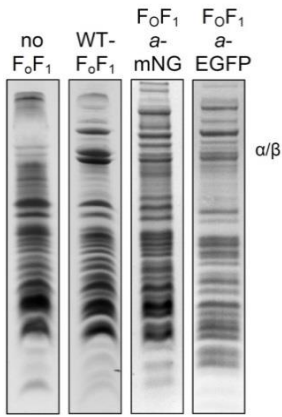

**Figure 2.** Relative expression levels of $F_oF_1$-ATP synthase mutants. Collected cell membranes were subjected to 12 % SDS-PAGE. After staining of the gel with coomassie, the α and β subunits of the 'wildtype' $F_oF_1$-ATP synthase (lane 2) could be seen clearly. Fusion proteins at the C terminus of the $a$-subunit resulted in a highly reduced relative expression level of $F_oF_1$-ATP synthases according to the intensities of α and β subunits (lane 3, $F_oF_1$-$a$-mNeonGreen, and lane 4, $F_oF_1$-$a$-EGFP). Note that lane 4 for $F_oF_1$-$a$-EGFP taken from[43].

Our purification procedures for ATP synthesis-active $F_oF_1$-ATP synthase from *Escherichia coli* were based on well-established protocols[5, 11, 31, 41]. However, the low expression level of the $F_oF_1$-$a$-mNeonGreen mutant made purification more difficult. Enzyme purification was divided in five steps: membrane preparation, solubilization and precipitation, intermediate size exclusion chromatography, ion exchange chromatography and final size exclusion chromatography (Figures 3 and 4). During membrane preparation, $F_oF_1$-ATP synthase was enriched because two washing steps removed several membrane-associated proteins from the membrane. These washing steps, in combination with the strain RA1 lacking the cytochrome *bo* quinol oxidase, were important for the purity of $F_oF_1$-ATP synthase. They reduced the amount of cytochrome-containing contaminants absorbing at 405 nm.

After membrane solubilization with DDM (Figure 4, lane Sol-S) solubilized membrane proteins were concentrated and lipids removed by ammonium sulfate precipitation.

Solubilized membrane proteins were subjected to a XK16/100 Sephacryl S300 column after resolving in size exclusion buffer (Figure 3A, Figure 4 lane S300). This step removed more lipids and separated $F_oF_1$-ATP synthase from small molecular weight impurities (likeley nucleotides and ammonium sulfate) which would interfere with the following anion exchange chromatography. The SDS-PAGE revealed that size exclusion chromatography further enriched the $F_oF_1$-ATP synthase (Figure 4 lane S300). Fractions 11 to 14 were subjected to the anion exchange chromatography (Figure 3A, light and dark grey), whereas only the ion exchange chromatogram of fraction 13 (Figure 3A, dark grey) is shown here (Figure 3B). Fractions 9 and 10 were not used, because they contained mainly protein aggregates (void volume). Furthermore, fractions 15 to 22 were also discarded, because they contained only small amounts of $F_oF_1$-ATP synthase but were colored brownish (most probably from lipids).

Following injection of the S300 fractions on the anion exchange chromatography column, a small fraction of the proteins did not bind to the strong anion exchanger and was eluted immediately (Figure 3B). The column was washed with several column volumes until the UV absorbance signal reached a baseline. With increasing concentration of KCl, the $F_oF_1$-ATP synthase eluted from the column with a maximum at around 400 mM KCl. Tightly bound membrane proteins eluted as a peak at around 900 mM KCl. The $F_oF_1$-ATP synthase peak exhibited a broad shoulder ranging from fraction 1 to 5 indicating impurities eluting together with the $F_oF_1$-ATP synthase. Therefore, only fractions 6 and 7 were combined ('pooled') and precipitated by ammonium sulfate. Only these protein fractions showed optical absorption at 506 nm. The corresponding SDS-PAGE (Figure 4 lane IEX) showed that the anion exchange step further separated the $F_oF_1$-ATP synthase from the majority of impurities.

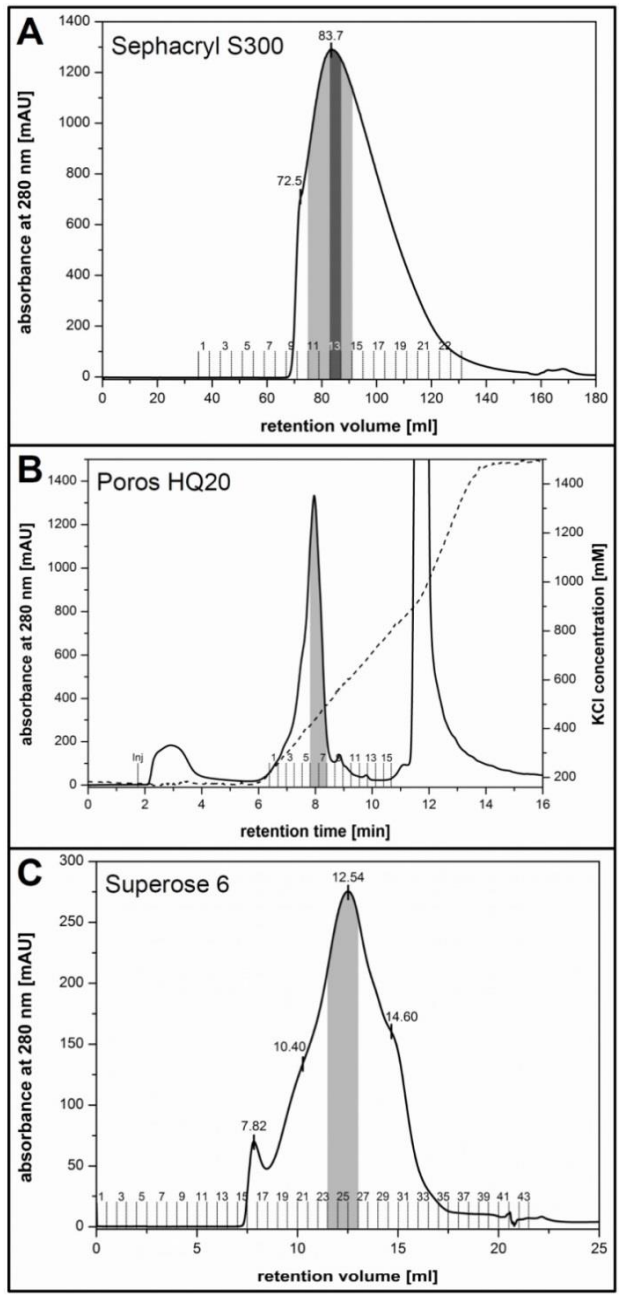

**Figure 3.** Chromatography steps for $F_oF_1$-*a*-mNeonGreen purification. (A) The solubilized membrane proteins were loaded on a XK16/100 Sephacryl S300 size exclusion column. The fractions with the highest ATP hydrolysis rates (light grey) were selected and loaded separately on a Poros HQ 20 anion exchange column. (B) The anion exchange run shown here was from fraction 13 (see A, highlighted in dark grey). Because of impurities eluting at the front of the $F_oF_1$-ATP synthase peak (a shoulder before fraction 5), only fractions 6 and 7 (light grey) were combined and loaded subsequently on a Superose 6 10/300 GL column. (C) Only fractions 24-26 (light grey) were used for further experiments. Peak retention volumes are shown in A and C, fraction numbers are indicated, as well as the point of injection in B.

The anion exchange fractions 6 and 7 of the S300-fractions 11 to 14 were pooled, concentrated and subjected to a final size exclusion chromatography for 'polishing'. The major modification to established protocols was the exchange of the column in this step. The previously used XK16/100 Sephacryl S400 column (geometric volume: 180 ml) was too large for the small amount of enzyme resulting from the low expression level. Furthermore, the Sephacryl S400 fractionation range is very wide (20 kDa – 8 MDa). Therefore we now used a Tricorn Superose 6 10/300 GL column (GE Healthcare, USA). With a geometric volume of only 25 ml and a narrower fractionation range (5 kDa – 5 MDa), the use of this column resulted in a sharper and more concentrated separation of the ATP synthase from impurities. Figure 3C shows the elution profile of the size exclusion. The major peak (retention volume 12.54 ml) ranging from fraction 24 to 26 corresponded to the $F_oF_1$-ATP synthase (Figure 4: Superose 6). The SDS-PAGE shows that all subunits are present. However, also some impurities are still visible. Only these fractions were used for further studies.

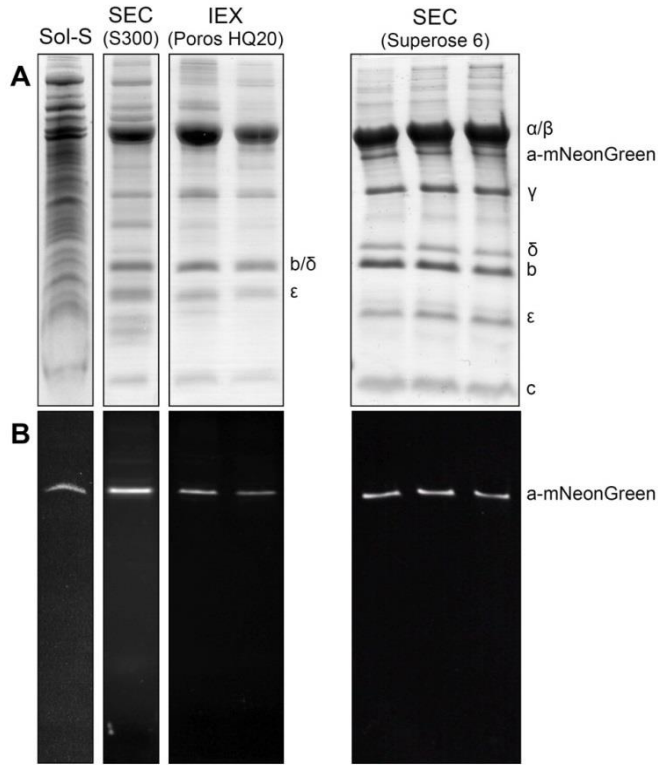

**Figure 4.** Purification of $F_oF_1$-*a*-mNeonGreen. Protein samples of the major purification steps were subjected to 12 % SDS-PAGE. Before staining with coomassie (A) a fluorogram of the in-gel fluorescence of mNeonGreen was made (B). After solubilization of membranes (Sol-S) only the α and β-subunits were visible. Other $F_oF_1$-ATP synthase subunits could not be distinguished from impurities. After the first size exclusion chromatography (S300, fraction 13, see Figure 3A), the other enzyme subunits were detectable. The majority of protein impurities were removed by the subsequent ion exchange step (IEX, fractions 6 and 7 in Figure 3B). The final size exclusion chromatography (Superose 6, fractions 24-26 in Figure 3C) resulted in an enzyme containing all subunits of $F_o$ (*a, b, c*) and of $F_1$ (α, β, γ, δ, ε). The fluorogram in (B) showed that the *a*-mNeonGreen fusion was folded correctly and incorporated into the $F_oF_1$ holoenzyme complex.

### 3.3 Fluorescence labeling of the *c* subunit in $F_oF_1$-*a*-mNeonGreen ATP synthase

To monitor the rotary motion of a *c* subunit in the $F_o$ motor by smFRET, mNeonGreen on the stator subunit *a* is the FRET donor. The FRET acceptor is bound to a cysteine at position 2 of the *c* subunit using cysteine-reactive dyes. We tested the labeling of the *c* subunit with three different dyes. The degree of $F_oF_1$-ATP synthase labeling was 59 % for AlexaFluor 647, 86 % for ATTO 647N and 109 % for AlexaFluor 568. Labeling efficiencies exceeding 100 % are possible because one $F_oF_1$-ATP synthase contains 10 *c* subunits. Labeling specificity was checked by SDS-PAGE. Fluorescence of mNeonGreen and the FRET acceptor were detected independently by two fluorograms that were overlaid in Figure 5. Labeling with AlexaFluor 647 and ATTO 647N were highly specific, but labeling with AlexaFluor 568 resulted in some nonspecific labeling of proteins with molecular weights around the subunit *a*-mNeonGreen fusion as well as subunits α and β.

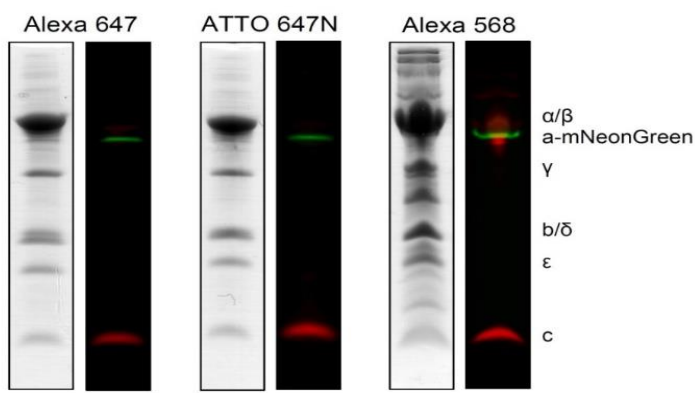

**Figure 5.** Fluorescence labeling of subunit *c* in $F_oF_1$-*a*-mNeonGreen on 12% SDS-PAGE. Fluorescence of mNeonGreen and of the respective label at the *c* subunit was detected subsequently with different excitation and emission filters. The fluorograms were false-colored and overlayed. Fluorescence of mNeonGreen is shown in green and of the FRET acceptor in red, respectively. The corresponding coomassie staining of subunits is shown at the left of each fluorogram.

## 3.4 Catalytic activities of $F_oF_1$-$a$-mNeonGreen ATP synthase

The activities of the mNeonGreen-tagged $F_oF_1$-ATP synthase were measured for both ATP hydrolysis and ATP synthesis. Note that ATP hydrolysis activity only gives limited information about the quality of the holoenzyme, because also a $F_1$ fragment is able to hydrolyze ATP with high rates. Using an ATP-regenerating assay which prevented a rapid product inhibition, the amount of hydrolyzed ATP was proportional to the absorbance decrease of NADH (Figure 6A). This decrease was significantly accelerated by addition of LDAO (a high activation ratio by addition of LDAO served as an indicator for the presence of the ε subunit in $F_oF_1$-ATP synthase[73, 74]). Figure 6B summarizes the relative ATP hydrolysis turnover rates of the 'wildtype' enzyme, $F_oF_1$-$a$-mNeonGreen with or without labeled cysteine in subunit $c$ and $F_oF_1$-$a$-EGFP. The EGFP and the mNeonGreen fusions without cysteine labeling exhibited the same hydrolysis rates as the 'wildtype' enzyme, indicating that the fusion to the $a$ subunit had no effect on ATP hydrolysis activity. Interestingly, AlexaFluor 647-labeled $F_oF_1$-$a$-mNeonGreen showed elevated ATP hydrolysis rates. The LDAO activation of all $F_oF_1$-$a$-mNeonGreen variants was higher than 'wildtype' $F_oF_1$-ATP synthase, ruling out an apparent higher ATP hydrolysis activity because of a dissociation or loss of subunit ε in these preparations.

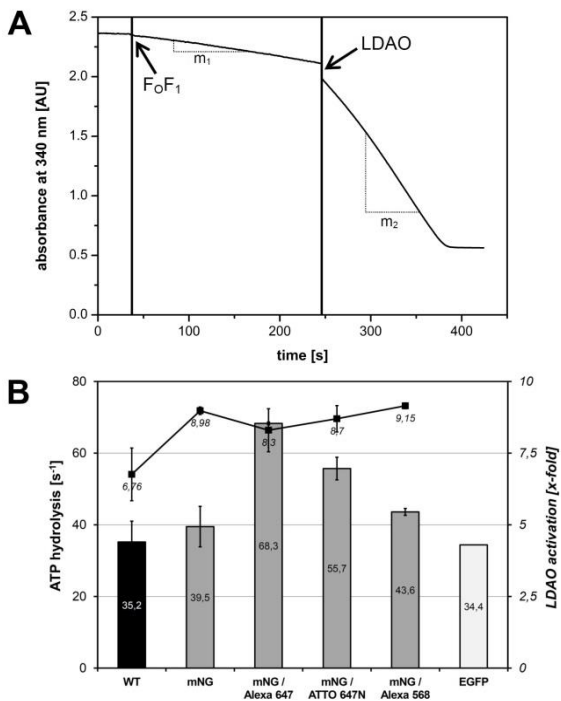

**Figure 6.** Measurement and comparison of ATP hydrolysis rates. (A) Time trace of a ATP hydrolysis measurement. Addition of $F_oF_1$-ATP synthase to the reaction buffer started ATP hydrolysis and resulted in a decrease of the absorbance of NADH at 340 nm (slope $m_1$). Subsequent addition of LDAO resulted in an activation of ATP hydrolysis (slope $m_2$). (B) ATP hydrolysis turnover in ATP molecules per $F_oF_1$-ATP synthase. Reference activity was set by the 'wildtype' (WT) for comparison with unlabeled $F_oF_1$-$a$-mNeonGreen ('mNG'), three differently labeled $F_oF_1$-$a$-mNeonGreen and $F_oF_1$-$a$-EGFP (EGFP). ATP hydrolysis rates (bars) and LDAO activation ratio (black line) of the mNeonGreen mutation with or without FRET acceptor dye were comparable to or slightly higher than the rates of the 'wildtype' enzyme. ATP hydrolysis rate of $F_oF_1$-$a$-EGFP was measured by M. G. Düser (PhD thesis, University of Freiburg, 2008).

The capability to synthesize ATP is the strongest measure of the quality of a $F_oF_1$-ATP synthase preparation, because for ATP synthesis the $F_oF_1$ holoenzyme has to be correctly folded, assembled and reconstituted into the membrane. Only for ATP synthesis, all rotor parts have to be functional, i.e. $F_1$ motor and $F_o$ motor have to be tightly coupled. Figure 7A shows the principle of the ATP synthesis measurement. First, proteoliposomes containing one $F_oF_1$-ATP synthase at maximum were pre-incubated in an acidic buffer. After injection of these proteoliposomes into the basic buffer, a proton gradient over the membrane was formed that started ATP synthesis. ATP was used by the luciferase to oxidize d-luciferin to oxyluciferin. This reaction emitted light at a wavelength of 560 nm that was detected by a photomultiplier. The high affinity of the luciferase to ATP ensured the high sensitivity of this assay. Figure 7B shows a typical time trace of an ATP synthesis measurement. Injection of proteoliposomes resulted in an immediate rise in luminescence. After a few seconds, the increase of luminescence slowed down and reached a maximum level due to dissipation of the proton gradient or proton-motive force across the membrane. The ATP synthesis rates were calculated from the initial slope by calibrating the background luminescence (which reflects the amount of residual ATP in the ADP solution used for the reaction buffer). Figure 7C summarizes the ATP synthesis turnover rates of 'wildtype' enzyme, of $F_oF_1$-$a$-mNeonGreen with or without labeled cysteine on the $c$ subunit and of $F_oF_1$-$a$-EGFP[17]. All subunit $a$ fusion mutants with or without labeled subunit $c$ showed turnover rates comparable to the 'wildtype' enzyme. AlexaFluor 568-labeled $F_oF_1$-$a$-

mNeonGreen exhibited an apparently higher turnover rate. However, for reconstitution of this enzyme, a different batch of pre-formed liposomes was used that could possibly result in different ATP synthesis rates.

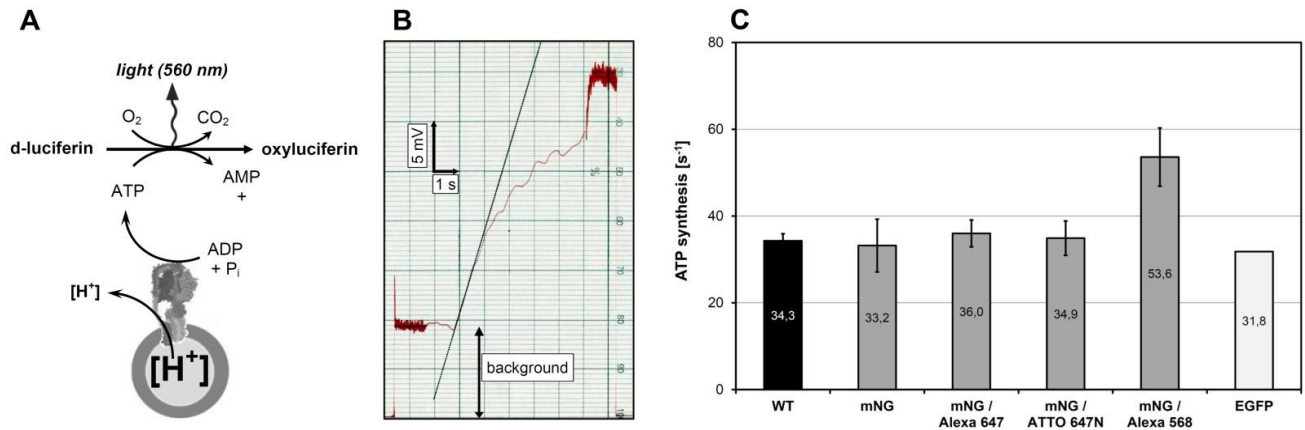

**Figure 7.** Measurement and comparison of ATP synthesis rates. (A) Scheme of the ATP synthesis measurement. Proteoliposomes were preincubated in acidic buffer and then injected into a basic buffer containing d-luciferin and luciferase. The ATP produced by the $F_oF_1$-ATP synthase was instantly used by the luciferin/luciferase system and emitted light. (B) Raw data of an ATP synthesis measurement. Rising luminescence intensities indicated ATP synthesis. After a few seconds, the proton gradient over the membrane dissipated and the ATP synthesis slows down. (C) ATP synthesis rates in ATP molecules per $F_oF_1$-ATP synthase and with 'wildtype' $F_oF_1$-ATP synthase (WT) being the reference, unlabeled $F_oF_1$-$a$-mNeonGreen (mNG), three differently labeled $F_oF_1$-$a$-mNeonGreen and $F_oF_1$-$a$-EGFP (EGFP). Neither the fusion of mNeonGreen to the C terminus of subunit $a$ nor the labeling of subunit $c$ altered ATP synthesis activities.

## 4. DISCUSSION

We described the generation, expression and purification of an active $F_oF_1$-ATP synthase containing a C-terminal fusion of the fluorescent protein mNeonGreen to subunit $a$ with additional E2C substitution at the $c$ subunit. This enzyme $F_oF_1$-$a$-mNeonGreen could be specifically labeled with cysteine-reactive dyes as FRET acceptors. In the future, this two-color labeled $F_o$ motor can be combined with differently labeled $F_1$ domains for three-color smFRET measurements, for example $F_oF_1$-ATP synthase comprising mNeonGreen as a primary FRET donor at the cytoplasmic side of the $a$ subunit, a secondary FRET donor at the central rotary stalk of the $F_1$ domain (i.e. the ε or γ subunit) and an FRET acceptor at the $c$ subunit. This new FRET donor mNeonGreen could eventually circumvent the limitations of our previous studies using EGFP as the primary FRET donor[16, 17, 21, 32, 33]. EGFP exhibited different photophysical artefacts on the single-molecule fluorescence level like spectral fluctuations and blinking.

The $F_oF_1$-$a$-mNeonGreen mutant showed a similar expression behavior as other enzyme mutants carrying a fusion protein at the C terminus of $a$. No differences in phenotype could be observed between 'wildtype' $F_oF_1$, $F_oF_1$-$a$-mNeonGreen and $F_oF_1$-$a$-EGFP. All three mutants could grow on succinate as sole carbon and energy source indicating that all three $F_oF_1$-ATP synthases are active *in vivo*. Interestingly, the SDS-PAGE of the membrane fractions of these three proteins revealed that the expression of both $F_oF_1$-$a$-mNeonGreen and $F_oF_1$-$a$-EGFP was significantly reduced (by 60-80 %). This behavior was found for other mutants as well: $F_oF_1$-ATP synthases with a Snap-tag or a Halo-tag at the C terminus of subunit $a$ led to the same loss of enzyme expression level, and similarly but less pronounced (40-50 % reduction) also chromosome-encoded fusions to the C terminus of subunit $a$ showed a lowered enzyme expression level (G. Deckers-Hebestreit, personal observation). These results are in accord with other mutations within the intergenic region between *atpB* and *atpE*. This intergenic region includes a translation initiation region (TIR)[75, 76] which is needed to produce 10 times more $c$ subunits than $a$ subunit with both genes localized on the same polycistronic mRNA strand. Mutations in close vicinity to the TIR often resulted in a massive reduction of subunit $c$ and, therefore, also of $F_oF_1$-ATP synthase (G. Deckers-Hebestreit, personal observation)[43]. However, the addition of two amino acids encoded by "GAAACA" before the stop codon of the $a$ subunit had no effect on the expression level. The fusion protein sequences were introduced exactly at the same position indicating that the mechanism of disturbing the translation initiation does not depend on the primary sequence alone, but also depends on the secondary structure far before the TIR side within the *a* subunit.

Despite the limitations resulting from the lower expression level of the $F_oF_1$-*a*-mNeonGreen mutant, we could successfully purify the $F_oF_1$-*a*-mNeonGreen ATP synthase. The quality of the enzyme was comparable to the purified $F_oF_1$-*a*-EGFP. However, the enzyme exhibited more impurities than purified 'wildtype' $F_oF_1$ ATP synthase[41]. The most important purification step of the latter protocol was the anion exchange chromatography (Figure 3B). Potentially this step could be improved because it is also responsible for the lower purity of the enzymes with fluorescent protein fusions. A closer look on the elution profile of the 'wildtype' protein revealed that the $F_oF_1$-ATP synthase peak profile also contained a shoulder in a comparable fraction. This indicated that during 'wildtype' or fusion mutant purifications, a protein impurity was co-eluting with similar proportion. However, the total amount of eluted $F_oF_1$-ATP synthase is much higher in the 'wildtype' preparation leading to a better enzyme-to-impurity ratio.

Finally, we have changed the column for the last 'polishing' chromatography step. The smaller Superose 6 column offered a more confined fractionation range that was directly visible during enzyme purification. The elution profile of the original Sephacryl S400 column consisted only of one peak with a very small shoulder. Therefore, this step was only a desalting step rather than a purification step. In contrast, the $F_oF_1$-ATP synthase eluted from the Superose 6 column as a central peak with wide shoulders to both sides (Figure 3C). Using only the central fractions of the eluate the proportion of protein impurities could be significantly reduced. Moreover, due to the small volume of the column, the eluates were better concentrated and the run was finished in only one-fifth of the time compared to the Sephacryl S400 column run which makes the column also attractive for highly expressed $F_oF_1$-ATP synthase mutants. The remaining protein impurities would only cause problems, if they interfere with the cysteine labeling or with enzyme reconstitution and ATP synthesis; a situation that was not observed for $F_oF_1$-*a*-mNeonGreen with cysteine-labeled subunit *c*.

For a FRET-labeled $F_oF_1$-ATP synthase, the *c* subunit had to be labeled comprising an introduced cysteine residue at position 2. We evaluated three different fluorophores with different chemical structures, charges and hydrophobicity. The AlexaFluor 647 $C_2$-maleimide represented an indocarbocyanine dye like Cy5. With a net charge of -3 this fluorophore is very hydrophilic. The ATTO 647N maleimide is a carborhodamine with a net charge of +1 and a moderate hydrophilic character. AlexaFluor 568 $C_5$-maleimide belongs to the class of rhodamines and is hydrophilic due to a net charge of -2. The labeling results showed that the binding of the fluorophore at the *c* subunit was achievable in each case with high efficiency and specificity. Only AlexaFluor 568 exhibited some nonspecific labeling, but fine-tuning of the degree of labeling is possible by variation of incubation times, temperature, pH, or the dye-to-protein ratio.

ATP synthesis and hydrolysis rates of the new mNeonGreen fusion to $F_oF_1$-ATP synthase (before and after FRET acceptor labeling of the *c* subunit) revealed that the $F_oF_1$-*a*-mNeonGreen mutant is fully active as compared to the 'wildtype' enzyme. Contrary to $F_oF_1$-*a*-EGFP showing a slightly reduced ATP synthesis activity after labeling with AlexaFluor 568 maleimide[21], the mNeonGreen fusion enzyme is also fully active after labeling of subunit *c*. Hence, we could generate a fully active $F_oF_1$ ATP synthase that has two FRET labels at the $F_o$ domain. After reconstitution of this FRET-labeled $F_oF_1$ ATP synthase and subsequent removal of the $F_1$ domain, a separately prepared and fluorescently labeled $F_1$ can be re-attached to yield a $F_oF_1$-ATP synthase with three fluorophores. We will use the triple-labeled enzyme for smFRET experiments to measure elastic deformations within the two coupled rotary motors of the $F_oF_1$ holoenzyme, with the aim to improve the temporal and spatial resolution of conformational dynamics measurements compared to our preliminary experiments[22, 31-33].

**Acknowlegdements**

We thank Brigitte Herkenhoff-Hesselmann (University of Osnabrück) and Sonja Rabe (Jena University Hospital) for excellent technical assistance. This work was supported in part by the Collaborative Research Center 944 'Physiology and dynamics of cellular microcompartments' (to G.D.-H.) and by the 'Incentive Award' of the Faculty of Biology/Chemistry of the University of Osnabrück (to G.D.-H.). Funding by DFG grant BO1891/15-1 (to M.B.) is gratefully acknowledged.

# REFERENCES

[1] Borsch, M., Turina, P., Eggeling, C., Fries, J. R., Seidel, C. A., Labahn, A. and Graber, P., "Conformational changes of the H+-ATPase from Escherichia coli upon nucleotide binding detected by single molecule fluorescence," *FEBS letters* 437, 251-254 (1998).
[2] Borsch, M., Diez, M., Zimmermann, B., Reuter, R. and Graber, P., "Stepwise rotation of the gamma-subunit of EF(0)F(1)-ATP synthase observed by intramolecular single-molecule fluorescence resonance energy transfer," *FEBS letters* 527, 147-152 (2002).
[3] Steinbrecher, T., Hucke, O., Steigmiller, S., Borsch, M. and Labahn, A., "Binding affinities and protein ligand complex geometries of nucleotides at the F(1) part of the mitochondrial ATP synthase obtained by ligand docking calculations," *FEBS letters* 530, 99-103 (2002).
[4] Borsch, M., Diez, M., Zimmermann, B., Trost, M., Steigmiller, S. and Graber, P., "Stepwise rotation of the gamma-subunit of EFoF1-ATP synthase during ATP synthesis: a single-molecule FRET approach," *Proc. SPIE* 4962, 11-21 (2003).
[5] Diez, M., Zimmermann, B., Borsch, M., Konig, M., Schweinberger, E., Steigmiller, S., Reuter, R., Felekyan, S., Kudryavtsev, V., Seidel, C. A. and Graber, P., "Proton-powered subunit rotation in single membrane-bound FoF1-ATP synthase," *Nature structural & molecular biology* 11, 135-141 (2004).
[6] Diez, M., Borsch, M., Zimmermann, B., Turina, P., Dunn, S. D. and Graber, P., "Binding of the b-subunit in the ATP synthase from Escherichia coli," *Biochemistry* 43, 1054-1064 (2004).
[7] Boldt, F. M., Heinze, J., Diez, M., Petersen, J. and Borsch, M., "Real-time pH microscopy down to the molecular level by combined scanning electrochemical microscopy/single-molecule fluorescence spectroscopy," *Anal Chem* 76, 3473-3481 (2004).
[8] Steigmiller, S., Zimmermann, B., Diez, M., Borsch, M. and Graber, P., "Binding of single nucleotides to H+-ATP synthases observed by fluorescence resonance energy transfer," *Bioelectrochemistry* 63, 79-85 (2004).
[9] Steigmiller, S., Borsch, M., Graber, P. and Huber, M., "Distances between the b-subunits in the tether domain of F(0)F(1)-ATP synthase from E. coli," *Biochimica et biophysica acta* 1708, 143-153 (2005).
[10] Zarrabi, N., Zimmermann, B., Diez, M., Graber, P., Wrachtrup, J. and Borsch, M., "Asymmetry of rotational catalysis of single membrane-bound F0F1-ATP synthase," *Proc. SPIE* 5699, 175-188 (2005).
[11] Zimmermann, B., Diez, M., Zarrabi, N., Graber, P. and Borsch, M., "Movements of the epsilon-subunit during catalysis and activation in single membrane-bound H(+)-ATP synthase," *Embo J* 24, 2053-2063 (2005).
[12] Krebstakies, T., Zimmermann, B., Graber, P., Altendorf, K., Borsch, M. and Greie, J. C., "Both rotor and stator subunits are necessary for efficient binding of F1 to F0 in functionally assembled Escherichia coli ATP synthase," *The Journal of biological chemistry* 280, 33338-33345 (2005).
[13] Borsch, M. and Graber, P., "Subunit movement in individual H+-ATP synthases during ATP synthesis and hydrolysis revealed by fluorescence resonance energy transfer," *Biochemical Society transactions* 33, 878-882 (2005).
[14] Zimmermann, B., Diez, M., Borsch, M. and Graber, P., "Subunit movements in membrane-integrated EF0F1 during ATP synthesis detected by single-molecule spectroscopy," *Biochimica et biophysica acta* 1757, 311-319 (2006).
[15] Galvez, E., Duser, M., Borsch, M., Wrachtrup, J. and Graber, P., "Quantum dots for single-pair fluorescence resonance energy transfer in membrane- integrated EFoF1," *Biochemical Society transactions* 36, 1017-1021 (2008).
[16] Duser, M. G., Zarrabi, N., Bi, Y., Zimmermann, B., Dunn, S. D. and Borsch, M., "3D-localization of the a-subunit in FoF1-ATP synthase by time resolved single-molecule FRET," *Proc. SPIE* 6092, 60920H (2006).
[17] Duser, M. G., Bi, Y., Zarrabi, N., Dunn, S. D. and Borsch, M., "The proton-translocating a subunit of F0F1-ATP synthase is allocated asymmetrically to the peripheral stalk," *The Journal of biological chemistry* 283, 33602-33610 (2008).
[18] Zarrabi, N., Duser, M. G., Ernst, S., Reuter, R., Glick, G. D., Dunn, S. D., Wrachtrup, J. and Borsch, M., "Monitoring the rotary motors of single FoF1-ATP synthase by synchronized multi channel TCSPC," *Proc. SPIE* 6771, 67710F (2007).
[19] Zarrabi, N., Duser, M. G., Reuter, R., Dunn, S. D., Wrachtrup, J. and Borsch, M., "Detecting substeps in the rotary motors of FoF1-ATP synthase by Hidden Markov Models," *Proc. SPIE* 6444, 64440E (2007).
[20] Johnson, K. M., Swenson, L., Opipari, A. W., Jr., Reuter, R., Zarrabi, N., Fierke, C. A., Borsch, M. and Glick, G. D., "Mechanistic basis for differential inhibition of the F(1)F(o)-ATPase by aurovertin," *Biopolymers* 91, 830-840 (2009).
[21] Duser, M. G., Zarrabi, N., Cipriano, D. J., Ernst, S., Glick, G. D., Dunn, S. D. and Borsch, M., "36 degrees step size of proton-driven c-ring rotation in FoF1-ATP synthase," *Embo J* 28, 2689-2696 (2009).
[22] Zarrabi, N., Ernst, S., Duser, M. G., Golovina-Leiker, A., Becker, W., Erdmann, R., Dunn, S. D. and Borsch, M., "Simultaneous monitoring of the two coupled motors of a single FoF1-ATP synthase by three-color FRET using duty cycle-optimized triple-ALEX," *Proc. SPIE* 7185, 718505 (2009).
[23] Borsch, M., Reuter, R., Balasubramanian, G., Erdmann, R., Jelezko, F. and Wrachtrup, J., "Fluorescent nanodiamonds for FRET-based monitoring of a single biological nanomotor F[sub o]F[sub 1]-ATP synthase," *Proc. SPIE* 7183, 71832N (2009).
[24] Borsch, M., "Targeting cytochrome C oxidase in mitochondria with Pt(II)-porphyrins for photodynamic therapy," *Proc. SPIE* 7551, 75510G (2010).
[25] Borsch, M., "Single-molecule fluorescence resonance energy transfer techniques on rotary ATP synthases," *Biological chemistry* 392, 135-142 (2011).
[26] Borsch, M. and Wrachtrup, J., "Improving FRET-based monitoring of single chemomechanical rotary motors at work," *Chemphyschem* 12, 542-553 (2011).
[27] Rendler, T., Renz, M., Hammann, E., Ernst, S., Zarrabi, N. and Borsch, M., "Monitoring single membrane protein dynamics in a liposome manipulated in solution by the ABELtrap," *Proc. SPIE* 7902, 79020M (2011).

[28] Seyfert, K., Oosaka, T., Yaginuma, H., Ernst, S., Noji, H., Iino, R. and Borsch, M., "Subunit rotation in a single F[sub o]F[sub 1]-ATP synthase in a living bacterium monitored by FRET," *Proc. SPIE* 7905, 79050K (2011).

[29] Renz, M., Rendler, T. and Borsch, M., "Diffusion properties of single FoF1-ATP synthases in a living bacterium unraveled by localization microscopy," *Proc. SPIE* 8225, 822513 (2012).

[30] Hammann, E., Zappe, A., Keis, S., Ernst, S., Matthies, D., Meier, T., Cook, G. M. and Borsch, M., "Step size of the rotary proton motor in single FoF1-ATP synthase from a thermoalkaliphilic bacterium by DCO-ALEX FRET," *Proc. SPIE* 8228, 82280A (2012).

[31] Ernst, S., Duser, M. G., Zarrabi, N., Dunn, S. D. and Borsch, M., "Elastic deformations of the rotary double motor of single FoF1-ATP synthases detected in real time by Förster resonance energy transfer," *Biochimica et Biophysica Acta (BBA) - Bioenergetics* 1817, 1722-1731 (2012).

[32] Ernst, S., Duser, M. G., Zarrabi, N. and Borsch, M., "Monitoring transient elastic energy storage within the rotary motors of single FoF1-ATP synthase by DCO-ALEX FRET," *Proc. SPIE* 8226, 82260I (2012).

[33] Ernst, S., Duser, M. G., Zarrabi, N. and Borsch, M., "Three-color Förster resonance energy transfer within single FoF1-ATP synthases: monitoring elastic deformations of the rotary double motor in real time," *J Biomed Opt* 17, 011004 (2012).

[34] Borsch, M., "Microscopy of single FoF1-ATP synthases— The unraveling of motors, gears, and controls," *IUBMB life* 65, 227-237 (2013).

[35] Sielaff, H. and Borsch, M., "Twisting and subunit rotation in single FOF1-ATP synthase," *Phil Trans R Soc B* 368, 20120024 (2013).

[36] Sielaff, H., Heitkamp, T., Zappe, A., Zarrabi, N. and Borsch, M., "Subunit rotation in single FRET-labeled F1-ATPase hold in solution by an anti-Brownian electrokinetic trap," *Proc. SPIE* 8590, 859008 (2013).

[37] Zarrabi, N., Clausen, C., Duser, M. G. and Borsch, M., "Manipulating freely diffusing single 20-nm particles in an Anti-Brownian Electrokinetic Trap (ABELtrap)," *Proc. SPIE* 8587, 85870L (2013).

[38] Borsch, M. and Duncan, T. M., "Spotlighting motors and controls of single FoF1-ATP synthase," *Biochemical Society transactions* 41, 1219-1226 (2013).

[39] Bockenhauer, S. D., Duncan, T. M., Moerner, W. E. and Borsch, M., "The regulatory switch of F1-ATPase studied by single-molecule FRET in the ABEL Trap," *Proc. SPIE* 8950, 89500H (2014).

[40] Zarrabi, N., Ernst, S., Verhalen, B., Wilkens, S. and Borsch, M., "Analyzing conformational dynamics of single P-glycoprotein transporters by Förster resonance energy transfer using hidden Markov models," *Methods* 66, 168-179 (2014).

[41] Heitkamp, T., Sielaff, H., Korn, A., Renz, M., Zarrabi, N. and Borsch, M., "Monitoring subunit rotation in single FRET-labeled FoF1-ATP synthase in an anti-Brownian electrokinetic trap," *Proc. SPIE* 8588, 85880Q (2013).

[42] Duncan, T. M., Duser, M. G., Heitkamp, T., McMillan, D. G. and Borsch, M., "Regulatory conformational changes of the epsilon subunit in single FRET-labeled FoF1-ATP synthase," *Proc. SPIE* 8948, 89481J (2014).

[43] Renz, A., Renz, M., Klutsch, D., Deckers-Hebestreit, G. and Borsch, M., "3D-localization microscopy and tracking of FoF1-ATP synthases in living bacteria," *Proc. SPIE* 9331, 93310D (2015).

[44] Su, B., Duser, M. G., Zarrabi, N., Heitkamp, T., Starke, I. and Borsch, M., "Observing conformations of single FoF1-ATP synthases in a fast anti-Brownian electrokinetic trap," *Proc. SPIE* 9329, 93290A (2015).

[45] Borsch, M., "Unraveling the rotary motors in FoF1-ATP synthase by single-molecule FRET," in *Advanced Time-Correlated Single Photon Counting Applications* Becker, W., Ed., pp. 309-338, Springer (2015).

[46] Deckers-Hebestreit, G. and Altendorf, K., "The F0F1-type ATP synthases of bacteria: structure and function of the F0 complex," *Annual review of microbiology* 50, 791-824 (1996).

[47] Etzold, C., Deckers-Hebestreit, G. and Altendorf, K., "Turnover number of Escherichia coli F0F1 ATP synthase for ATP synthesis in membrane vesicles," *European journal of biochemistry / FEBS* 243, 336-343 (1997).

[48] Birkenhager, R., Greie, J. C., Altendorf, K. and Deckers-Hebestreit, G., "F0 complex of the Escherichia coli ATP synthase. Not all monomers of the subunit c oligomer are involved in F1 interaction," *European journal of biochemistry / FEBS* 264, 385-396 (1999).

[49] Altendorf, K., Stalz, W., Greie, J. and Deckers-Hebestreit, G., "Structure and function of the F(o) complex of the ATP synthase from Escherichia coli," *The Journal of experimental biology* 203, 19-28 (2000).

[50] Greie, J. C., Deckers-Hebestreit, G. and Altendorf, K., "Secondary structure composition of reconstituted subunit b of the Escherichia coli ATP synthase," *European journal of biochemistry / FEBS* 267, 3040-3048 (2000).

[51] Greie, J. C., Deckers-Hebestreit, G. and Altendorf, K., "Subunit organization of the stator part of the F0 complex from Escherichia coli ATP synthase," *Journal of bioenergetics and biomembranes* 32, 357-364 (2000).

[52] Deckers-Hebestreit, G., Greie, J., Stalz, W. and Altendorf, K., "The ATP synthase of Escherichia coli: structure and function of F(0) subunits," *Biochimica et biophysica acta* 1458, 364-373 (2000).

[53] Stalz, W. D., Greie, J. C., Deckers-Hebestreit, G. and Altendorf, K., "Direct interaction of subunits a and b of the F0 complex of Escherichia coli ATP synthase by forming an ab2 subcomplex," *The Journal of biological chemistry* 278, 27068-27071 (2003).

[54] Krebstakies, T., Aldag, I., Altendorf, K., Greie, J. C. and Deckers-Hebestreit, G., "The stoichiometry of subunit c of Escherichia coli ATP synthase is independent of its rate of synthesis," *Biochemistry* 47, 6907-6916 (2008).

[55] Ballhausen, B., Altendorf, K. and Deckers-Hebestreit, G., "Constant c10 ring stoichiometry in the Escherichia coli ATP synthase analyzed by cross-linking," *J Bacteriol* 191, 2400-2404 (2009).

[56] Brandt, K., Muller, D. B., Hoffmann, J., Hubert, C., Brutschy, B., Deckers-Hebestreit, G. and Muller, V., "Functional production of the Na+ F1F(O) ATP synthase from Acetobacterium woodii in Escherichia coli requires the native AtpI," *Journal of bioenergetics and biomembranes* 45, 15-23 (2013).

[57] Busch, K. B., Deckers-Hebestreit, G., Hanke, G. T. and Mulkidjanian, A. Y., "Dynamics of bioenergetic microcompartments," *Biological chemistry* 394, 163-188 (2013).

[58] Brockmann, B., Koop Genannt Hoppmann, K. D., Strahl, H. and Deckers-Hebestreit, G., "Time-delayed in vivo assembly of subunit a into preformed Escherichia coli FoF1 ATP synthase," *J Bacteriol* 195, 4074-4084 (2013).

[59] Brandt, K., Maiwald, S., Herkenhoff-Hesselmann, B., Gnirss, K., Greie, J. C., Dunn, S. D. and Deckers-Hebestreit, G., "Individual interactions of the b subunits within the stator of the Escherichia coli ATP synthase," *The Journal of biological chemistry* 288, 24465-24479 (2013).

[60] Hilbers, F., Eggers, R., Pradela, K., Friedrich, K., Herkenhoff-Hesselmann, B., Becker, E. and Deckers-Hebestreit, G., "Subunit delta is the key player for assembly of the H(+)-translocating unit of Escherichia coli F(O)F1 ATP synthase," *The Journal of biological chemistry* 288, 25880-25894 (2013).

[61] Deckers-Hebestreit, G., "Assembly of the Escherichia coli FoF1 ATP synthase involves distinct subcomplex formation," *Biochemical Society transactions* 41, 1288-1293 (2013).

[62] Shaner, N. C., Lambert, G. G., Chammas, A., Ni, Y., Cranfill, P. J., Baird, M. A., Sell, B. R., Allen, J. R., Day, R. N., Israelsson, M., Davidson, M. W. and Wang, J., "A bright monomeric green fluorescent protein derived from Branchiostoma lanceolatum," *Nature methods* 10, 407-409 (2013).

[63] Moriyama, Y., Iwamoto, A., Hanada, H., Maeda, M. and Futai, M., "One-step purification of Escherichia coli H(+)-ATPase (F0F1) and its reconstitution into liposomes with neurotransmitter transporters," *The Journal of biological chemistry* 266, 22141-22146 (1991).

[64] Kuo, P. H., Ketchum, C. J. and Nakamoto, R. K., "Stability and functionality of cysteine-less F(0)F1 ATP synthase from Escherichia coli," *FEBS letters* 426, 217-220 (1998).

[65] Aggeler, R., Ogilvie, I. and Capaldi, R. A., "Rotation of a gamma-epsilon subunit domain in the Escherichia coli F1F0-ATP synthase complex. The gamma-epsilon subunits are essentially randomly distributed relative to the alpha3beta3delta domain in the intact complex," *The Journal of biological chemistry* 272, 19621-19624 (1997).

[66] Fischer, S., Etzold, C., Turina, P., Deckers-Hebestreit, G., Altendorf, K. and Graber, P., "ATP synthesis catalyzed by the ATP synthase of Escherichia coli reconstituted into liposomes," *European journal of biochemistry / FEBS* 225, 167-172 (1994).

[67] Fischer, S. and Graber, P., "Comparison of DeltapH- and Delta***φ***-driven ATP synthesis catalyzed by the H(+)-ATPases from Escherichia coli or chloroplasts reconstituted into liposomes," *FEBS letters* 457, 327-332 (1999).

[68] Holloway, P. W., "A simple procedure for removal of Triton X-100 from protein samples," *Analytical biochemistry* 53, 304-308 (1973).

[69] Rosing, J., Harris, D. A., Kemp, A., Jr. and Slater, E. C., "Nucleotide-binding properties of native and cold-treated mitochondrial ATPase," *Biochimica et biophysica acta* 376, 13-26 (1975).

[70] Fischer, S., Graber, P. and Turina, P., "The activity of the ATP synthase from Escherichia coli is regulated by the transmembrane proton motive force," *The Journal of biological chemistry* 275, 30157-30162 (2000).

[71] Schagger, H. and von Jagow, G., "Tricine-sodium dodecyl sulfate-polyacrylamide gel electrophoresis for the separation of proteins in the range from 1 to 100 kDa," *Analytical biochemistry* 166, 368-379 (1987).

[72] Klionsky, D. J., Brusilow, W. S. and Simoni, R. D., "In vivo evidence for the role of the epsilon subunit as an inhibitor of the proton-translocating ATPase of Escherichia coli," *J Bacteriol* 160, 1055-1060 (1984).

[73] Lotscher, H. R., deJong, C. and Capaldi, R. A., "Interconversion of high and low adenosinetriphosphatase activity forms of Escherichia coli F1 by the detergent lauryldimethylamine oxide," *Biochemistry* 23, 4140-4143 (1984).

[74] Dunn, S. D., Tozer, R. G. and Zadorozny, V. D., "Activation of Escherichia coli F1-ATPase by lauryldimethylamine oxide and ethylene glycol: relationship of ATPase activity to the interaction of the epsilon and beta subunits," *Biochemistry* 29, 4335-4340 (1990).

[75] Schauder, B. and McCarthy, J. E., "The role of bases upstream of the Shine-Dalgarno region and in the coding sequence in the control of gene expression in Escherichia coli: translation and stability of mRNAs in vivo," *Gene* 78, 59-72 (1989).

[76] Mccarthy, J. E. G., Schairer, H. U. and Sebald, W., "Translational Initiation Frequency of Atp Genes from Escherichia-Coli - Identification of an Intercistronic Sequence That Enhances Translation," *Embo Journal* 4, 519-526 (1985).